\newif\ifsingle

\ifsingle
\documentclass[11pt,draftclsnofoot, onecolumn]{IEEEtran}		
\else		
\documentclass[10pt,final, twocolumn]{IEEEtran}
\fi

\usepackage{times}
\usepackage{amsmath,dsfont}
\usepackage{amssymb,amsthm}
\usepackage{epsfig,verbatim}
\usepackage{caption}
\usepackage{subcaption}
\usepackage{setspace}
\usepackage{color}
\usepackage{cite}
\usepackage{epstopdf}
\usepackage{graphicx}
\usepackage{accents}
\usepackage{acronym}
\usepackage[bookmarks,colorlinks]{hyperref}
\usepackage{booktabs}
\usepackage{mathtools} 
\usepackage{enumitem}
\usepackage{multirow}
\usepackage{soul}

\ifsingle

\newcommand{\figSpace}{\vspace{-0.2cm}}

\else

\newcommand{\figSpace}{\vspace{-0.3cm}}
\fi 
\acrodef{adc}[ADC]{analog-to-digital convertor}
\acrodef{cs}[CS]{compressed sensing}
\acrodef{cnn}[CNN]{convolutional neural network} 
\acrodef{dnn}[DNN]{deep neural network} 
\acrodef{csi}[CSI]{channel state information}
\acrodef{map}[MAP]{maximum a-posteriori probability}
\acrodef{snr}[SNR]{signal-to-noise ratio}
\acrodef{sinr}[SINR]{signal-to-interference-and-noise ratio}
\acrodef{bs}[BS]{base station} 
\acrodef{iot}[IoT]{Internet of Things}
\acrodef{mimo}[MIMO]{multiple-input multiple-output}
\acrodef{mse}[MSE]{mean-squared error}
\acrodef{pdf}[PDF]{probability density function}
\acrodef{rv}[RV]{random variable}
\acrodef{fec}[FEC]{forward error correction} 
\acrodef{lti}[LTI]{linear time-invariant}
\acrodef{wss}[WSS]{wide-sense stationary}
\acrodef{psd}[PSD]{power spectral density}
\acrodef{ser}[SER]{symbol error rate} 
\acrodef{ber}[BER]{bit error rate} 
\acrodef{sgd}[SGD]{stochastic gradient descent} 
\acrodef{isi}[ISI]{intersymbol interference}  
\acrodef{awgn}[AWGN]{additive white Gaussian noise} 
\acrodef{ut}[UT]{user terminal} 
\acrodef{mmw}[mmWave]{millimeter wave}
\acrodef{ai}[AI]{artificial intelligence} 
\acrodef{dma}[DMA]{dynamic metasurface antenna}
\acrodef{pga}[PGA]{projected gradient ascent}
\acrodef{quadriga}[QuaDRiGa]{Quasi Deterministic Radio channel Generator}

\IEEEoverridecommandlockouts

\title{AI-Empowered Hybrid MIMO Beamforming}

\author{  
	\IEEEauthorblockN{
         Nir Shlezinger,~\IEEEmembership{Member,~IEEE}, 
         Mengyuan Ma, \IEEEmembership{Student Member, IEEE},
         Ortal Lavi,~\IEEEmembership{Student Member,~IEEE},\\
         Nhan Thanh Nguyen,~\IEEEmembership{Member, IEEE},
        Yonina C. Eldar, \IEEEmembership{Fellow, IEEE}, 
        and Markku Juntti, \IEEEmembership{Fellow, IEEE}
 \\
	}   
} 
 
\begin{document}
	
	\maketitle
 	\pagestyle{plain}  
\thispagestyle{plain}

\begin{abstract}
Hybrid \ac{mimo} is an attractive technology for realizing  extreme  massive \ac{mimo} systems envisioned for future wireless communications in a scalable and power-efficient manner. However, the fact that hybrid \ac{mimo} systems implement part of their beamforming in analog and part in digital makes the optimization of their beampattern notably more challenging compared with conventional fully digital \ac{mimo}. Consequently, recent years have witnessed a growing interest in using data-aided \ac{ai} tools for hybrid beamforming design. 
This article reviews candidate strategies to leverage data to improve real-time hybrid beamforming design. We discuss the architectural constraints and characterize the core challenges associated with hybrid beamforming optimization. We then present how these challenges are treated via conventional optimization, and identify different \ac{ai}-aided design approaches. These can be roughly divided into purely data-driven deep learning models and different forms of deep unfolding techniques for combining \ac{ai} with classical optimization. We provide a systematic comparative study between  existing approaches including both numerical evaluations and qualitative measures. We conclude by presenting future research opportunities associated with the incorporation of \ac{ai} in hybrid \ac{mimo} systems. 
\end{abstract}

\section{Introduction} 
Massive \ac{mimo} systems and high frequency communications at millimeter wave (mmWave) and sub-Thz bands are expected to play a key role in future sixth-generation (6G) networks~\cite{giordani2020toward}. These technologies are naturally supportive of each other, as massive \ac{mimo} using large transmit and receive antenna arrays facilitates generating highly focused beams that are essential for reliable communications at high frequencies, while short wavelength signaling enables packing \ac{mimo} configurations with a massive number of elements in a limited aperture. 
However, implementing such massive \ac{mimo} transceivers gives rise to several challenges. One of these core challenges is associated with the notable cost and power consumption of RF chains  operating at high frequencies, which in conventional {\em fully digital} \ac{mimo} arrays separately connect each antenna element to the digital signal processing unit. 


{\em Hybrid beamforming} is considered to be a leading solution for coping with the above challenge,  enabling high frequency massive \ac{mimo} communications with a limited number of RF chains~\cite{molisch2017hybrid}. This is achieved by delegating part of the signal processing to the analog domain, thus, dividing the beamforming task into digital and analog counterparts. The possible beampatterns achievable in analog are dictated by the circuitry, with typical implementations based on phase shifters~\cite{yu2016alternating}, vector modulators~\cite{zirtiloglu2022power}, and \acp{dma}~\cite{shlezinger2021dynamic}. Consequently, hybrid transceivers are inherently constrained in their beamforming capabilities compared with fully digital ones.

While hybrid designs alleviate some of the cost and power issues of massive \ac{mimo} systems, their constrained form gives rise to algorithmic and signal processing challenges. Most notably, the beamforming task, i.e., the translation of \ac{csi} into a suitable beampattern, involves solving a typically non-convex constrained optimization problem. Various iterative optimization algorithms have been proposed for tuning hybrid beamformers~\cite{qiao2020alternating}, differing in their considered hardware constraints and objective. A key limitation of these iterative solutions stems from their typically slow convergence, as the beampattern setting must be done in real-time to cope with channel variations. 

The emergence of deep learning as an enabler technology for \ac{ai} has led to the proposal of \ac{ai}-empowered hybrid beamfoming designs. While deep learning typically deals with setting an inference rule based on data, one can also train \acp{dnn} to tackle challenging optimization problems~\cite{zappone2019wireless}. Once trained, \acp{dnn} infer at fixed latency, dictated by the number of layers, and can thus be used to rapidly map \ac{csi} into beampatterns~\cite{elbir2019hybrid}. An alternative approach to leverage data for hybrid beamforming arises  from {\em model-based deep learning} methodologies~\cite{shlezinger2022model}. Here, deep learning techniques are used to enhance iterative hybrid beamforming optimizers rather than replacing them, while data is exploited to achieve rapid convergence~\cite{agiv2022learn,nguyen2023deep, balevi2021unfolded}. The proliferation of different approaches for hybrid \ac{mimo} beamforming motivates a unified overview of these methods. 

In this article, we provide a systematic tutorial of \ac{ai}-aided methodologies for hybrid \ac{mimo} beamforming. While successfully realizing hybrid \ac{mimo} transceivers inevitably combines hardware developments with signal processing algorithmic considerations, we focus  on the latter, without restricting our attention to a specific  implementation. 
We start by discussing hybrid \ac{mimo} systems, reviewing  representative architectures and describing how their operation impacts the achievable beampatterns. We  pinpoint the design challenges arising from hybrid beamforming, and identify the aspects that motivate incorporating \ac{ai}.   

Next, we describe hybrid beamforming design approaches, dividing them into three main families: {\em Optimization-based} methods, which employ iterative optimizers for setting the beampatterns; {\em \ac{dnn}-based} schemes, where  \ac{csi} is mapped into hybrid configurations via a pre-trained \ac{dnn}; and {\em Deep-unfolded} designs, where deep learning techniques are leveraged to facilitate iterative optimization. For the latter, we identify different types of  unfolding approaches, and discuss how each gives rise to a different design. Based on this division, we provide a comparative study, including both a numerical study and a qualitative comparison, where we identify the interplay between the approaches in terms of several key figures-of-merit. We conclude by discussing research challenges that are left for future exploration, and are expected to  pave the way towards harnessing the potential of \ac{ai} for hybrid \ac{mimo} systems.

\begin{figure*}
    \centering
    \includegraphics[width=\linewidth]{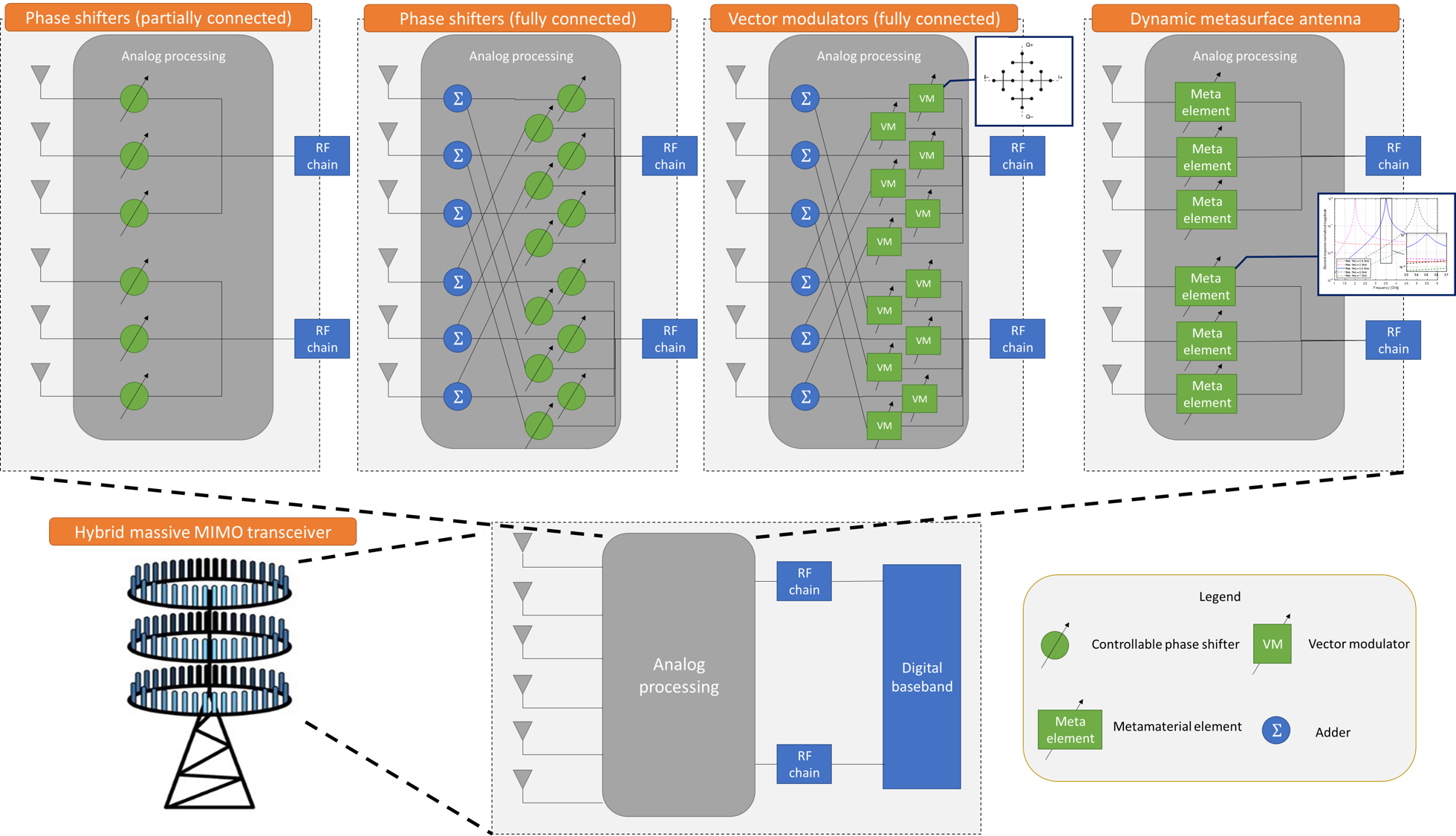}
    \caption{Schematic illustration of different hybrid \ac{mimo} transceiver architectures and their corresponding analog processing model, including partially and fully connected phase shifter networks, vector modulators, and \acp{dma}.}
    \label{fig:Hybrid_Illust1}
\end{figure*}
   
\section{Hybrid Beamforming}
\label{sec:Transceivers}

\subsection{Hybrid  \ac{mimo} Transceivers}
\label{subsec:Hybrid}
Massive \ac{mimo} transceivers are equipped with an antenna array comprised of a large number of elements, denoted $M$. In the current 5G \acp{bs}, $M$ can be on the order of several tens. This number is expected to grow to possibly thousands of antennas in 6G, when evolving from massive \ac{mimo} to holographic \ac{mimo}~\cite{huang2020holographic}. In  conventional fully digital \ac{mimo} architectures, the signal being fed to each antenna is processed separately digitally, by having a digital processing unit connect to each antenna via a dedicated RF chain. 

In hybrid \ac{mimo} systems, the number of RF chains, denoted $K$, is smaller than that of antennas. This is achieved via analog processing that interfaces the RF chains with the antennas, as illustrated in Fig.~\ref{fig:Hybrid_Illust1}. The analog processor 
achieves different manipulations of the signals. 
A natural benefit of hybrid \ac{mimo} over fully digital architectures stems from the fact that it uses less RF chains than antenna elements, which becomes a crucial factor when using large scale arrays in high frequencies. 
In addition to reducing RF chains, hybrid designs can also facilitate interference rejection as well as mitigate  distortion induced by low-resolution \aclp{adc}~\cite{zirtiloglu2022power}.

\subsection{Architectural Constraints}
\label{subsec:Constraints}
Hybrid \ac{mimo} systems combine digital and analog signal processing. The processing part carried out in the digital domain is highly flexible, allowing to effectively apply different mappings to different spectral components. However, analog processing is highly constrained, and the set of different mappings it can realize is dictated by its hardware, with several different hardware architectures proposed in the literature. To exemplify the constraints associated with different designs, we briefly review a few representative analog architectures, focusing on their operation in transmission:

\subsubsection{Phase Shifter Networks} 
The most commonly considered analog hardware employs phase shifters 
with controllable phases~\cite{molisch2017hybrid}. These are typically divided into fully connected architectures, where a dedicated phase shifter connects each RF chain with each antenna, and to partially connected structures, in which each RF chain is connected to a single antenna via a dedicated phase shifter. Often in practice, the phases applied by each phase shifter cannot be arbitrarily set, and must comply to some predefined phase resolution. Furthermore, phase shifters are typically designed to (approximately) preserve the same phase shift over a considered band. Thus, they are often modelled as applying the same mapping to each spectral component. 

\subsubsection{Discrete Vector Modulators} While phase shifters  only affect the phase of the signal, vector modulators are analog circuits that can realize different combinations of phase shifting and signal attenuation. Such forms of analog circuitry provide additional flexibility compared with phase shifters, due to its ability to also affect the magnitude of the signals in a controllable fashion. Nonetheless, low-power vector modulators are typically constrained to realize only a predefined finite number of different phase-attenuation combinations~\cite{zirtiloglu2022power}. 

\subsubsection{Dynamic Metasurface Antennas} 
An emerging technology for realizing holographic \ac{mimo} is to utilize metasurfaces, that are planar configurations of controllable metamaterial elements, as antennas.
Unlike the aforementioned architectures, which rely on the incorporation of dedicated analog circuitry, \acp{dma} implement configurable analog processing as an inherent byproduct of their antenna structure~\cite{shlezinger2021dynamic}. When transmitting, the signal at the output of each RF chain propagates along a waveguide, and is radiated from the elements connected to that waveguide, where each element can realize a form of a frequency-selective Lorentzian filter. Consequently, \acp{dma} inherently implement frequency-selective analog signal processing, which is constrained to take the Lorentzian form.

\subsection{Hybrid Beamforming Design Challenges}
\label{subsec:Challenges}
{\em Hybrid beamforming design}  is concerned with the joint setting of the analog and digital processing to optimize a predefined communication metric for the current channel realization. Typical metrics are the achievable rate or the minimal \ac{sinr} in multi-user communications. Focusing on  downlink transmission  with the common setting of linear beamforming, the task boils down to designing the precoders applied to the outgoing symbols in digital (where each spectral component can be precoded separately), along with the configuration of the analog processing. 

Hybrid beamforming design is associated with multiple core challenges, including:
\begin{enumerate}[label={\em C\arabic*},leftmargin=*]
    \item \label{itm:nonconvex} The resulting optimization problem based on which the digital precoders and the analog configuration are determined is rarely convex. Even when the design objective takes a quadratic form, e.g., the achievable rate of a linear Gaussian channel, the need to divide the processing into digital and analog parts, as well as the hardware constraints imposed on the analog processing, typically results in non-convex optimization. 
    \item \label{itm:rapid} Since hybrid beamforming is designed for a given channel realization, it needs to be carried out each time the channel conditions change, i.e., on each coherence duration (which can be as small as $125$ $\mu{\rm Sec}$ by 3GPP Release 17). As the coherence duration of wireless communication channels typically decreases with  carrier frequency, the design procedure must be performed rapidly to enable reliable communications within each coherence duration.
    \item \label{itm:robust} Hybrid beamforming design uses \ac{csi}, which is typically obtained from pilot signalling, and is thus likely to be noisy. Consequently, hybrid beamforming design should be able to cope with some level of error in its available \ac{csi}. 
\end{enumerate}

The above challenges, and particularly \ref{itm:nonconvex} and \ref{itm:rapid}, motivate \ac{ai}-aided designs, as discussed in the following section.

\section{\ac{ai}-Aided Hybrid Beamforming Design}
\label{sec:Beamforming}
\begin{figure*}
    \centering
    \includegraphics[width=\linewidth]{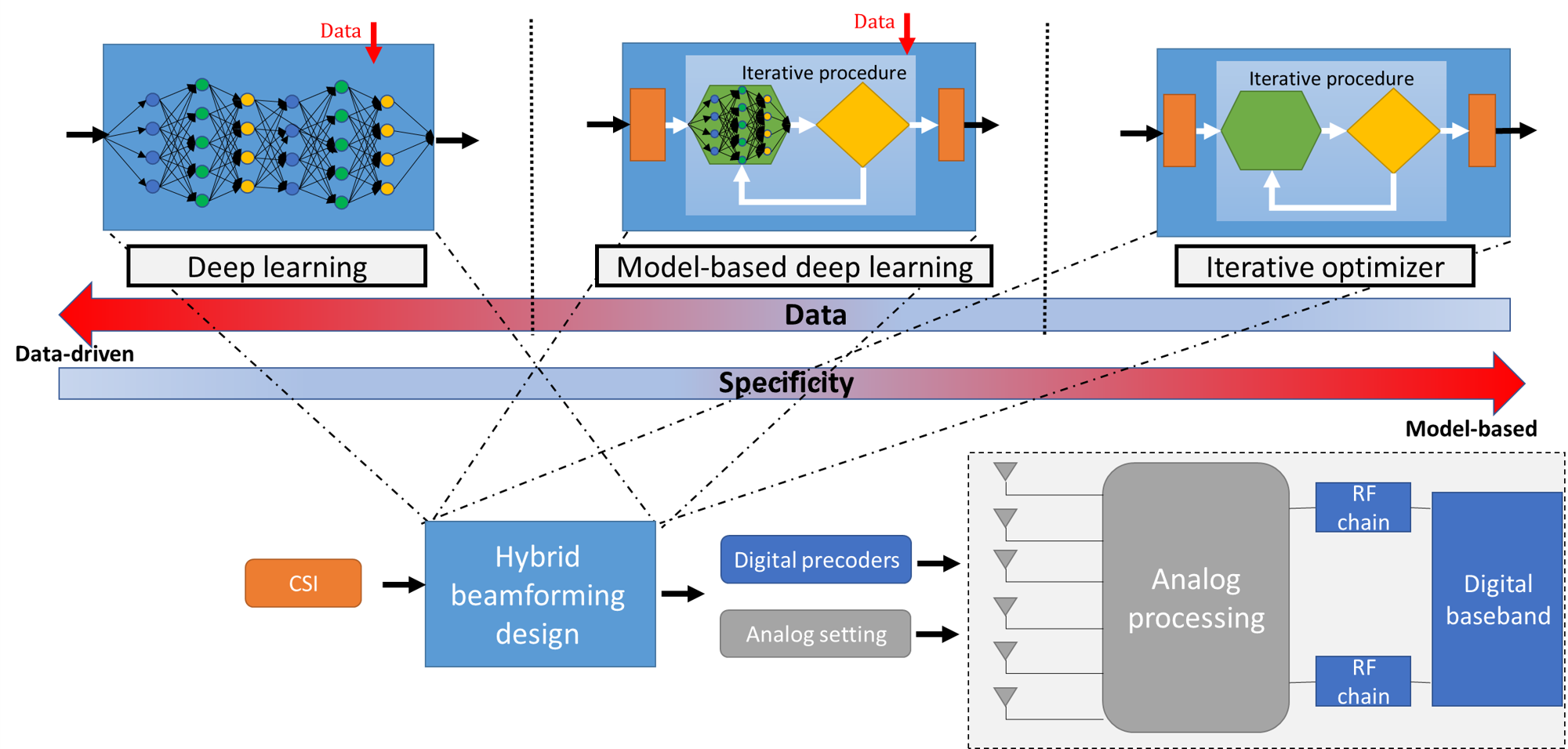}
    \caption{Illustration of different approaches for hybrid beamforming design.}
    \label{fig:Spectrum_1}
\end{figure*}

We next detail  leading frameworks for designing hybrid precoders. The first utilizes iterative optimizers that are specific to the problem at hand. The second employs \acp{dnn}, i.e., abstract architectures that are tuned  from data to map \ac{csi} into a hybrid beamformer configuration. The last  framework utilizes deep unfolding, which combines iterative optimization with deep learning via different forms of model-based deep learning~\cite{shlezinger2022model}. The latter constitutes a middle ground between the first two techniques by balancing specificity and data-driven learning capabilities, as illustrated in Fig.~\ref{fig:Spectrum_1}.

\subsection{Optimization-Based Hybrid Beamforming}
\label{subsec:Optimization}
As explained earlier, hybrid beamforming design is inherently an optimization problem. As such, it is traditionally tackled using optimization tools,  commonly via iterative solvers. Broadly speaking, there are two main approaches to cope with the non-convexity  (\ref{itm:nonconvex}):
\begin{itemize}
    \item A leading approach applies convex relaxation, i.e., formulates an alternative problem which is convex. Most commonly, the non-convex sum-rate objective in the hybrid configuration is  often replaced with seeking the hybrid setting that best approximates the fully digital rate-maximizing precoder~\cite{yu2016alternating, sohrabi2016hybrid}. Compared to directly maximizing the sum-rate, the relaxed formulation is often simpler to tackle, typically using iterative methods based on alternating optimization.  Yet, it may still result in a non-convex formulation, depending on the hardware constraints. 
    The resulting solution can be shown to approach the rate-maximizing setting in some regimes, and particularly when the number of RF chains $K$ is not smaller than the number of receive antennas~\cite{yu2016alternating}. 
    \item An alternative approach directly tackles the non-convex objective, typically by aiming to identify a suitable initial setting of the precoders and refine it using local-convex optimization techniques, e.g., \ac{pga}~\cite{agiv2022learn}. 
\end{itemize}

While iterative optimizers can often recover useful hybrid beamformers, they tend to require a large number of iterations to converge. As iterations are  translated into delay and  complexity, this property  limits their applicability in time-varying settings by \ref{itm:rapid}. While optimization theory provides techniques for reducing the number of iterations via, e.g., backtracking, such techniques involve additional lengthy computations during inference. 

\subsection{\ac{dnn}-Based Hybrid Beamforming}
\label{subsec:BlackBox}
Deep learning provides tools for tuning machine learning models parameterized as \acp{dnn} to learn a desirable complex mapping from data. \acp{dnn} can also be trained to tackle challenging optimization problems, such as those encountered in hybrid beamforming design~\cite{zappone2019wireless}. Architectures such as \acp{cnn} were shown to be capable of learning to map \ac{mimo} \ac{csi} into  analog and digital precoders~\cite{elbir2019hybrid}. 

\ac{dnn}-based inference rules are typically designed in a supervised manner, i.e., by providing data comprised of inputs and their desired outputs which the model learns to produce during training. However, for hybrid beamforming, they can often be trained {\em unsupervised}, namely, by providing a dataset comprised solely of channel realizations, without  specifying the desired beamformer for each channel. This is possible because the optimization objective, e.g., sum-rate or \ac{sinr}, can be evaluated for each selected precoders, while being differentiable with respect to them. Consequently, one can apply conventional gradient-based learning to training \ac{dnn}-based hybrid beamformers using the (negative) optimization objective as an unsupervised training loss~\cite{agiv2022learn}.    

\acp{dnn} are often computationally complex, being comprised of a large number of parameters, and their training can be lengthy. Yet, their latency during inference  is fixed based on the number of layers, and various software and hardware tools facilitate their parallelization. Consequently, using pre-trained \acp{dnn} for hybrid beamforming design is often more rapid compared with iterative optimizers. However, the usage of generic highly-parameterized models trained from data to replace optimization solvers gives rise to several drawbacks. First, the training of \acp{dnn} is often a lengthy  task, requiring large volumes of data (i.e., channel realizations) and tedious experimentation to  learn a suitable mapping. Furthermore, while their inference latency is fixed, the complexity of applying \acp{dnn} in terms of, e.g., floating point operations, is typically large compared with iterative optimizers, being dictated by the number of parameters. Moreover, \acp{dnn} are far less flexible compared with optimization methods, and each modification in the task, e.g., the incorporation of an additional user to the network, requires time-consuming retraining. Finally, \acp{dnn} are hardly  interpretable, in the sense that one can typically assign operational meaning only to their input and their output, and are typically treated as black boxes.

\subsection{Deep Unfolded Hybrid Beamforming}
\label{subsec:Unfolded}
Both mathematically principled iterative optimizers and data-driven deep learning systems have their individual limitations in the context of hybrid beamforming, motivating hybrid designs which leverage the individual strengths of each approach. Model-based deep learning~\cite{shlezinger2022model} accommodates a family of strategies for  combining inference based on principled mathematical models with deep learning techniques, with the methodology of {\em deep unfolding} being highly suitable for tasks typically tackled using iterative methods.

Deep unfolding is based on the similarity between the sequential operation of an iterative optimizer with $L$ iterations and the forward path of a \ac{dnn} with $L$ layers. Its underlying rationale treats the iterations as an inductive bias of a parameterized machine learning model,  effectively converting the optimizer with $L$ iterations, and thus with fixed latency, into a trainable discriminative model. This gives rise to three different forms of deep unfolded optimizers~\cite{shlezinger2022model}:
\begin{enumerate}[label={\em U\arabic*},leftmargin=*]
    \item \label{itm:hyper} {\em Learned Hyperparameters:} Iterative optimizers have hyperparameters, i.e., parameters of the solver, such as step sizes. The specific setting of these hyperparameters typically has little influence on the outcome of the optimizer when allowed to run until convergence, and are often tuned manually. However, when the optimizer is constrained to a fixed (and small) number of iterations, the hyperparameters setting can have a paramount effect. Deep unfolding can thus convert the hyperparameters of the iterative solver into trainable parameters, thus leveraging data to automatically tune iteration-specific hyperparameters within a predefined number of iterations. 
    \item \label{itm:obj} {\em Learned Objective:} Iterative optimizers are designed based on an objective function, e.g., sum-rate, and typically operate by modifying the optimization variable on each iteration to further improve the objective value. Deep unfolding can parameterize the objective function used in each iteration. This allows learning from data to have each iteration tune its optimization variable based on a different objective, such that the output after $L$ iterations would be most suitable in the sense of the true objective. 
    \item \label{itm:dnn} {\em \ac{dnn} Conversion:} The third form of deep unfolding designs a \ac{dnn} to imitate the operation of the iterative optimizer. This is typically achieved by replacing some of the operations in each iteration with trainable layers. Such unfolding supports different levels of abstractness. One can preserve the operation of the iterative optimizer while replacing only specific computations with trainable neurons, or alternatively design a highly-parameterized \ac{dnn} whose architecture is inspired by the operation of the optimizer from which it originates.
\end{enumerate} 

For hybrid beamforming, deep unfolded optimizers share the ability of \acp{dnn} to train in an unsupervised manner. Furthermore, the similarity of the architecture of unfolded optimizers to that of iterative optimizers brings forth additional factors which can facilitate training. First, the iterative optimizer can constitute a principled initialization for the trainable architecture, guaranteeing that training commences from a valid operation which intuitively should only be further improved as training progresses.  Moreover, the fact that the output of each trainable iteration can be associated with the optimization variable implies that the training can compute its loss not solely based on the output after $L$ iterations/layers, as in conventional \acp{dnn}, and can also account for the intermediate features. Such training losses, which are not applicable in black-box architectures, encourage the trainable model to produce valid settings at each iteration/layer, and thus constitute a regularization known to facilitate learning. 

The above methodologies, and particularly \ref{itm:hyper} (e.g., \cite{agiv2022learn,nguyen2023deep}) and \ref{itm:dnn} (e.g., \cite{balevi2021unfolded}),  enable data-aided fixed latency iterative optimization for hybrid beamforming design. In particular, deep unfolding with learned hyperparameters fully preserves the operation of the iterative optimizer, thus maintaining its flexibility and interpretability. Nonetheless, by learning different step size values for each iteration~\cite{agiv2022learn}, and even per optimized precoder values~\cite{nguyen2023deep}, one can notably reduce latency. Furthermore, the learned hyperparameters can be incorporated into optimizers applied with different objectives, including robust optimization for coping with \ac{csi} uncertainty \ref{itm:robust}, as shown in \cite{agiv2022learn}.


\section{Comparative Study}
\label{sec:RoadAhead}
In this section we compare the hybrid beamforming design approaches detailed earlier. To that aim, we present a numerical study comparing representative schemes from each design approach, after which we provide a qualitative comparison. 

\subsection{Numerical Evaluation}
\label{subsec:numerical}
To compare the considered hybrid beamforming design approaches, we simulate hybrid \ac{mimo} systems with fully-connected phase shifter network for analog processing. We compare the following methods for determining the precoders:
\begin{itemize}
    \item For optimization-based methods, we evaluate the  Riemannian manifold optimizer of \cite{yu2016alternating} and the alternating optimizer of \cite{sohrabi2016hybrid}, which are both based on convex relaxation of the sum-rate objective.
    \item For \ac{dnn}-based designs, we use a \ac{cnn} following the architecture of~\cite{elbir2019hybrid}, referred to as {\em black-box \ac{cnn}}. This architecture is comprised of three convolutional layers (with $3\times 3$ kernel) followed by three full-connected layers. The \ac{cnn} was trained to produce both the analog and digital precoders, as well to produce only the analog precoder while the digital precoder was tuned accordingly to best match the fully digital beamformer. As both implementations yielded similar results, only the latter is reported here.
    \item For unfolded optimizers, we consider both the ManNet model of \cite{nguyen2023deep}, that unfolds the convex-relaxed optimization, as well as the unfolded \ac{pga} of \cite{agiv2022learn}, which augments simple \ac{pga} steps applied to the non-convex sum-rate objective. Both these unfolded methods use merely 10 iterations while preserving the operation of the iterative optimizers from which they originate following \ref{itm:hyper}.
    \item To represent an upper bound on the achievable sum-rate, we  evaluate that achieved using fully digital beamforming. 
\end{itemize}

\begin{figure}
    \centering
    \includegraphics[width=\columnwidth]{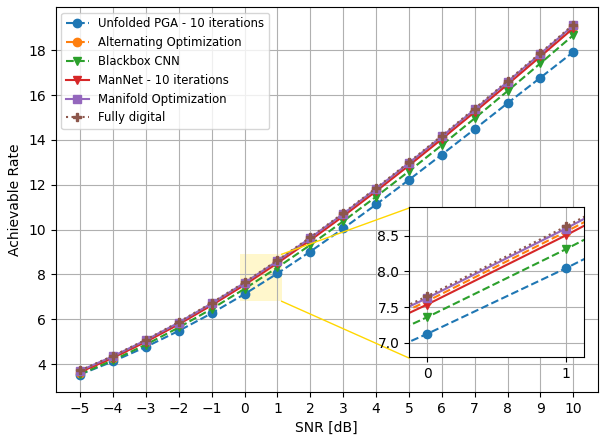}
    \figSpace
    \caption{Sum-rate vs. \acs{snr}, $4$ RF chains.}
    \label{fig: SNR}
\end{figure}

The considered \ac{mimo} transmitter has $M=12$ antennas, and serves $4$ single-antenna users by signalling over $16$ frequency bins. We generated $1000$ mmWave  channels with central frequency of $30$ GHz using the \acs{quadriga} model. 

\begin{figure}
    \centering
    \includegraphics[width=\columnwidth]{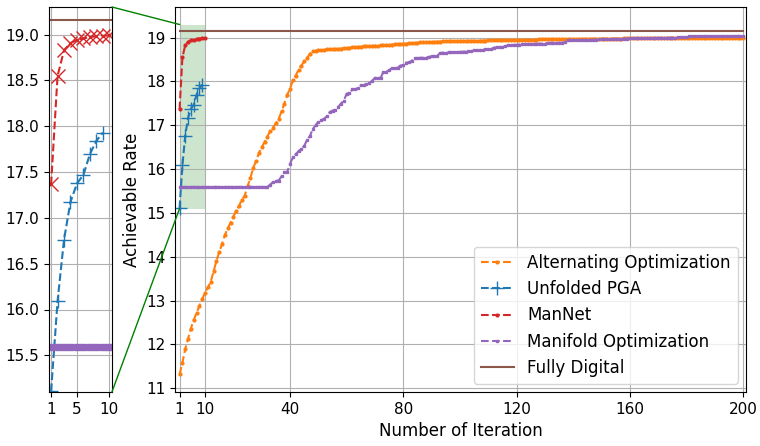}
    \figSpace
    \caption{Sum-rate per iteration.} 
    \label{fig: iter}
\end{figure}

We first set the number of RF chains to $K=4$, i.e., the same as the number of users. The resulting sum-rates versus \ac{snr}, depicted in Fig.~\ref{fig: SNR}, demonstrate that all optimizers based on convex relaxation, i.e., the iterative optimizers of \cite{yu2016alternating, sohrabi2016hybrid} and the \ac{ai}-aided ManNet~\cite{nguyen2023deep},  approach the sum-rate of fully digital beamforming. 
The black-box \ac{cnn} and the unfolded \ac{pga} are both within a small gap from the rate achieved with fully digital beamforming. Nonetheless, the gains of the unfolded designs over purely optimization-based methods is revealed when observing the number of iterations needed to achieve this performance. The sum-rate versus iteration for each iterative method at \ac{snr} of $10$ dB is reported in Fig.~\ref{fig: iter}. There, we observe that the unfolded methods leverage data to achieve their suitable settings with much less iterations compared with conventional iterative optimizers, indicating the ability of \ac{ai}-aided designs in notably reducing latency and computational complexity.

\begin{figure}
    \centering
    \includegraphics[width=\columnwidth]{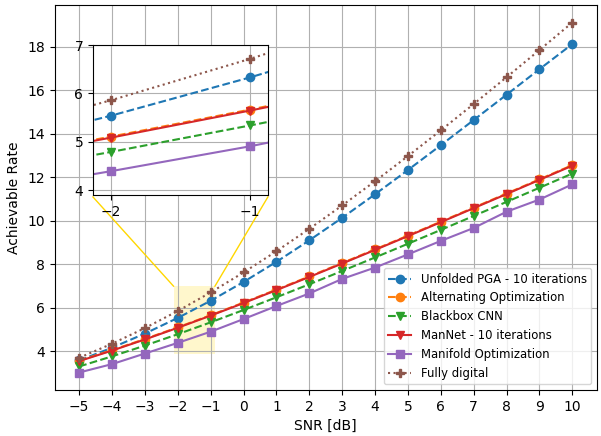}
    \figSpace
    \caption{Sum-rate vs. \ac{snr}, $2$ RF chains.}
    \label{fig: SNR_2}
\end{figure}

Another performance gain of \ac{ai}-aided designs over model-based optimizers is their ability to learn from data to cope with non-convexity. To see this, we repeat the  study of Fig.~\ref{fig: SNR} while setting the number of RF chains to $K=2$, i.e., less than the number of users, indicating a challenging regime for hybrid beamforming. The results, reported in Fig.~\ref{fig: SNR_2}, demonstrate that here the unfolded \ac{pga}, that directly tackles the non-convex sum-rate objective while leveraging data to learn to optimize, 
remains within a small gap of the fully digital upper bound. Here, the optimizers based on convex relaxation are notably outperformed, as they are designed to approach the fully digital beamformer, which cannot be achieved in this setting. 

\begin{table*}
\centering
{
\begin{tabular}{|p{2.2cm}|p{2.6cm}|p{2.6cm}|p{2.6cm}|p{2.6cm}|p{2.6cm}|}
 \hline
                              {\bf Method}                                    & {\bf Latency}    & {\bf Complexity}                                               & {\bf Data}            & {\bf Flexibility}       & {\bf Interpretability}                                                                                                      \\
                                \hline \hline 
Iterative Optimizers                               & High - numerous iterations  & Low - few operations in numerous iterations                                    & {\em None} - no data needed                                          & {\em Fully flexible} - applicable with different configurations                              & {\em Fully interpretable}                                          \\ \hline 
                                \acp{dnn}           & Medium - fixed by forward pass of \ac{dnn}       & High - complex high parameterized models                                  & High - massive data sets needed for training              & None - retraining is needed to switch configuration & Not interpretable \\   \hline 
                                Deep Unfolded Optimizers  \ref{itm:hyper}                    & {\em Lowest} - few predefined iterations of low complexity  & {\em Lowest} - few operations in few iterations       & Low - few parameters trained with small data sets &  {Flexible} - applicable with different configurations though performance may be affected & {\em Fully interpretable} - preserve operation as iterative optimizers \\  \hline
                                Deep Unfolded Optimizers  \ref{itm:dnn}                    & Low - few predefined iterations with moderate complexity  & Medium - complex parameterized mappings in few iterations          & Medium - relatively large number of parameters to train & None - retraining typically is needed to switch configuration & Partially interpretable as one can track intermediate features \\  \hline
\end{tabular}
}
    \caption{Qualitative comparison between the considered approaches for hybrid beamforming. }
    \label{Tbl:Comparison}
\end{table*}

\subsection{Qualitative Comparison}
\label{subsec:Comparison}
The approaches detailed earlier for optimizing hybrid beamformers differ in their properties, and are each suitable for different types of scenarios. The above numerical study allows to compare the approaches in terms of achievable rate, i.e., tackling \ref{itm:nonconvex}.   To shed light on additional meaningful comparative aspects, we next discuss five key figures-of-merit -- design latency, computational complexity, data requirements, flexibility, and interpretability. 
The comparison detailed below is summarized in Table~\ref{Tbl:Comparison}.

\subsubsection{Latency} A core challenge in hybrid beamforming is the need to update the beampattern on each coherence duration~\ref{itm:rapid}. Conventional iterative optimizers are typically lengthy, inducing notable latency due to their multiple iterations. This can be mitigated via deep unfolding, particularly via hyperparameter learning \ref{itm:hyper}, as demonstrated in Fig.~\ref{fig: iter}. Using \acp{dnn} for hybrid beamforming design typically has low latency, as computing the forward pass of a neural network with several layers is of fixed delay, which is reduced with parallelization and hardware accelerators, though not necessarily to the order of the coherence duration of wireless channels. 

\subsubsection{Complexity} While \acp{dnn} often support rapid and fixed-latency hybrid beamforming design, they are computationally complex, being comprised of a large number of parameters, and their limited latency is typically due to parallelization and hardware acceleration. Iterative optimizers are of a much smaller complexity, as each iteration typically involves a small number of operations, yet this complexity is not translated into low latency due to their  sequential operation. Deep unfolded designs,  particularly  with learned hyperparameters~\ref{itm:hyper}, share both the low complexity of iterative optimizers while supporting rapid inference due to their inherently fixed number of iterations. 

\subsubsection{Data} \ac{ai}-aided hybrid beamforming design leverages data to learn how to map \ac{csi} into hybrid precoders. While such learning can be done in an unsupervised manner, training \acp{dnn} for such tasks still requires large volumes of data, i.e., channel realizations from the same distribution as that expected at deployment. Deep unfolding balances the dependence on data by imposing an inductive bias on the learned model, trading parameterization for specificity~\cite{shlezinger2022model}, with abstract parameterizations (\ref{itm:dnn}) requiring more data compared with lesser parameterized models (\ref{itm:hyper}). 

\subsubsection{Flexibility} Hybrid beamforming design requires some level of flexibility, as channel configuration, e.g., the number of users, can change over time.  Iterative optimizers are extremely flexible, and the same optimizer can be applied in different settings. Similarly, unfolded methods that fully preserve the iterative optimizer (\ref{itm:hyper}) operation also share this flexibility. However, \acp{dnn} are trained for a fix configuration, and are thus highly non-flexible as they have to be retrained when the configuration changes. 

\subsubsection{Interpretability} An important property of hybrid beamforming design is the ability to understand how it maps the \ac{csi} into a hybrid precoder, and to track its processing chain. 
Iterative optimizers are fully interpretable, and so are unfolded optimizers which do not alter their operation (\ref{itm:hyper}). More abstract forms of unfolding that deviate from the optimizer  (\ref{itm:dnn}) are less interpretable, yet one can still track their procedure as  each iteration is still associated with an operational meaning. For black-box \acp{dnn} only the input and output have an interpretable value.

\section{Summary and Future Research Directions}
\label{sec:directions}
\ac{ai}-aided design and model-based deep learning bear the potential of notably facilitating real-time high-throughput hybrid beamforming, which in turn can pave the way towards sustainable and scalable massive \ac{mimo} deployments. However, several research directions are  to be explored to fully realize the potential of \ac{ai}-aided beamforming. We next review some candidate  topics.

\subsection{Hybrid \ac{mimo} with Integrated Sensing}
6G networks are envisioned to utilize \ac{mimo} transceivers not solely for communications, but also for sensing. Such operation induces various considerations on beamforming design, ranging from coexistence between sensing and communicating spectrum-sharing devices to dual-function signalling. These considerations notably complicate the setting of hybrid beamforming, as the optimization procedure has to account for additional aspects associated with the sensing functionality. This further motivates the exploration of \ac{ai}-aided techniques for hybrid \ac{mimo} with integrated sensing.

\subsection{Power and Hardware Oriented Designs}
While the majority of studies on hybrid \ac{mimo} consider phase shifter based analog circuitry, there are in fact various forms of hybrid architectures, each giving rise to different constraints affecting beamforming design. Furthermore, existing hybrid beamforming methods often overlook the fact that different configurations of the analog circuitry consume different powers. For instance, the ability to turn off a subset of the vector modulators in hybrid designs was shown to notably reduce power consumption~\cite{zirtiloglu2022power}. This motivates the exploration of hybrid beamforming algorithms that incorporate power and hardware considerations into their optimization procedure, and the associated excessive complexity motivates the usage of the advocated \ac{ai}-aided strategies. Moreover,  typical hybrid beamforming designs assume that the  antenna array is ideally linear, while practical  power amplifiers are nonlinear to varying degree. Linearization techniques are well known, but the overall array response design requires quite elaborate solutions. 

\subsection{Distributed Hybrid \ac{mimo}}
Future wireless communications are expected to deviate from conventional cellular architectures, utilizing multi-connectivity and cell-free topologies \cite{giordani2020toward}. This operation extends conventional centralized beamforming into distributed beamforming using a deployment of multiple collaborative \ac{mimo} transmitters. The reduced cost of hybrid architectures makes them suitable candidates for massive deployments, while the collaborative operation can overcome the limitations associated with their constrained beampatterns. The usage of \ac{ai} in such cases can notably facilitate real-time collaborative hybrid beamforming setting, possibly exploiting distributed machine learning paradigms such as federated learning and multi-agent reinforcement learning.

\subsection{From Far-Field to Near-Field} 
An additional consideration impacting beamforming in future wireless communications is the expected transition from far-field communications to near-field. In particular, the expected growth in the aperture of \ac{mimo} transceivers combined with the utilization of high frequencies implies that communications are likely to take place in the radiative near-field, as opposed to the conventional far-field assumed in traditional wireless transceiver designs. 

The operation in the radiative near-field brings forth new forms of beamforming, and in particular the ability to generate focused beams that can notably mitigate interference. Initial studies have unveiled that focused beams can also be achieved with different forms of hybrid beamforming using lengthy optimization~\cite{zhang2022beam}. Future studies are left to explore the ability to simultaneously support far-field and near-field users, and the ability of \ac{ai}-aided hybrid beamforming in enabling real-time and accurate forming of focused beampatterns for near-field communications. Furthermore, the spherical wavefronts of the near-field can in principle improve  the accuracy of positioning and other sensing applications, for which deep unfolded optimization is also a potential tool.

\bibliographystyle{IEEEtran}
\bibliography{IEEEabrv,refs}


\begin{IEEEbiographynophoto}{Nir Shlezinger} (nirshl@bgu.ac.il) is an Assistant Professor in the School of Electrical and Computer Engineering in Ben-Gurion University, Israel. 
\end{IEEEbiographynophoto}
\vskip -2\baselineskip plus -1fil

\begin{IEEEbiographynophoto}{Mengyuan Ma} (mengyuan.ma@oulu.fi) is a Ph.D. student in the Centre for Wireless Communications, University of Oulu, Finland.
\end{IEEEbiographynophoto}	
\vskip -2\baselineskip plus -1fill

\begin{IEEEbiographynophoto}{Ortal Lavi} (agivo@post.bgu.ac.il) is a graduate student in the School of Electrical and Computer Engineering in Ben-Gurion University, Israel.
\end{IEEEbiographynophoto}	
\vskip -2\baselineskip plus -1fill

\begin{IEEEbiographynophoto}{Nhan Thanh Nguyen} (nhan.nguyen@oulu.fi) received the Ph.D. degree from Seoul
National University of Science and Technology. He is currently
with University of Oulu, Finland.
\end{IEEEbiographynophoto}
\vskip -2\baselineskip plus -1fill
	
\begin{IEEEbiographynophoto}{Yonina C. Eldar} (yonina.eldar@weizmann.ac.il)
is a Professor in the Department of Math and Computer Science, Weizmann Institute of Science, Israel, where she heads the center for Biomedical Engineering and Signal Processing. She is a member of the Israel Academy of Sciences and Humanities, an IEEE Fellow and a EURASIP Fellow.
\end{IEEEbiographynophoto}	
\vskip -2\baselineskip plus -1fill

\begin{IEEEbiographynophoto}{Markku Juntti} (markku.juntti@oulu.fi) received the Dr.Sc. degree from
University of Oulu, Finland, where he has been a professor since 2000 and is the Head of CWC – Radio Technologies Research Unit. He is also an Adjunct Professor with the Department of Electrical and Computer Engineering, Rice University.  
\end{IEEEbiographynophoto}	
\vskip -2\baselineskip plus -1fill
	
\end{document}